\documentclass[12pt]{article}
\usepackage[english,german,french,polish]{babel}
\usepackage[T1]{fontenc}
\usepackage{amsfonts}

\begin{document}

\selectlanguage{english}

\baselineskip 0.75cm
\topmargin -0.6in
\oddsidemargin -0.1in

\let\ni=\noindent

\renewcommand{\thefootnote}{\fnsymbol{footnote}}

\pagestyle {plain}

\setcounter{page}{1}

\pagestyle{empty}

~~~

\begin{flushright}
IFT--05/13
\end{flushright}

{\large\centerline{\bf The simplest 3+2 model}}

{\large\centerline{\bf with two  light sterile neutrinos}}

\vspace{0.4cm}

{\centerline {\sc Wojciech Kr\'{o}likowski}}

\vspace{0.3cm}

{\centerline {\it Institute of Theoretical Physics, Warsaw University}}

{\centerline {\it Ho\.{z}a 69,~~PL--00--681 Warszawa, ~Poland}}

\vspace{0.6cm}

{\centerline{\bf Abstract}}

\vspace{0.2cm}

The simplest 3+2 neutrino model is described, where two light sterile neutrinos mix very weakly 
with three active neutrinos and mutually do not mix at all, while the mass-squared scale of the 
possible LSND effect is provided by $m_5^2 - m^2_4$ ($\nu_4$ and $\nu_5$ being two additional 
mass neutrinos connected with the existence of two sterile neutrinos $\nu_s$ and $\nu_{s'}$). This 
3+2 model is not better for explaining the LSND effect than the simplest 3+1 neutrino model, where 
one light sterile neutrino mixes very weakly with three active neutrinos, while the mass-squared 
scale of the possible LSND effect is given by $m_4^2 - m^2_1$ ($\nu_4$ denoting an additional 
mass neutrino existing due to a sterile neutrino $\nu_s$). However, a small LSND effect with 
amplitude of the order O($10^{-3}$) is not excluded by the present (pre-MiniBooNE) data.

\vspace{0.2cm}

\ni PACS numbers: 12.15.Ff , 14.60.Pq , 12.15.Hh .

\vspace{0.6cm}

\ni June 2005 

\vfill\eject

~~~
\pagestyle {plain}

\setcounter{page}{1}

\vspace{0.2cm}

As is well known, the neutrino mixing matrix $U^{(3)} = \left(U^{(3)}_{\alpha i} \right)
 \;(\alpha = e, \mu, \tau\;{\rm and}\; i=1, 2, 3)$ appearing in the unitary transformation

\begin{equation}
\nu_\alpha  = \sum_i U^{(3)}_{\alpha i}\, \nu_i 
\end{equation}

\ni  between the flavor neutrinos $\nu_e \,,\, \nu_\mu \,,\, \nu_\tau $ and mass neutrinos $\nu_1, 
\nu_2 , \nu_3 $  is experimentally consistent with the global bilarge form

\vspace{-0.1cm}

\begin{equation}
U^{(3)} = \left( \begin{array}{ccc} c_{12} & s_{12} & 0 
\\ - \frac{1}{\sqrt2} s_{12} & \frac{1}{\sqrt2} c_{12} & \frac{1}{\sqrt2}  \\ \frac{1}{\sqrt2} s_{12} 
& -\frac{1}{\sqrt2} c_{12} & \frac{1}{\sqrt2}  \end{array} \right) \;,
\end{equation}

\ni where $s^2_{12} \sim 0.30 $ [1], while $U^{(3)}_{e3} = s_{13} \exp(-i\delta)$ is negligible according to the negative result of Chooz experiment [2] (the upper limit is  $s^2_{13} < 0.03$). When neglecting 
$s_{13}$, the neutrino oscillation probabilities (in the vacuum) are

\begin{equation}
P(\nu_\alpha \rightarrow \nu_\beta) = \delta_{\beta \alpha} - 4\sum_{j>i} U^{(3)}_{\beta j} U^{(3)}_{\alpha j}\, U^{(3)}_{\beta i} U^{(3)}_{\alpha i} \sin^2 x_{j i}\,,
\end{equation}

\ni where 

\begin{equation} 
x_{j i} \equiv 1.27 \frac{\Delta m^2_{j i} L}{E}\; , \; \Delta m^2_{j i} \equiv m^2_j - m^2_i
\end{equation}

\ni  ($\Delta m^2_{j i}$, $L$ and $E$ are measured in eV$^2$, km and GeV, respectively), giving
experimentally $\Delta m^2_{21}\sim 8.0\times 10^{-5}{\rm eV}^2$ as well as $|\Delta m^2_{32}| 
\sim 2.2\times 10^{-3}{\rm eV}^2$ [1] for solar $\nu_e$'s and KamLAND reactor $\bar{\nu}_e$'s (with the solar MSW effect included) as well as atmospheric  and K2K accelerator $\nu_\mu$'s, respectively. 

The formulae (2) and (3) are consistent with the zero LSND effect. The nonzero LSND effect [3] would require the existence of a third neutrino mass-squared splitting, absent in the case of three active neutrinos only (unless the CPT invariance of neutrino oscillations is seriously violated, what does not seem to be realistic). The LSND effect will be tested soon in the ongoing MiniBooNE experiment [4]. If this test confirms the LSND effect, we will need at least one light sterile neutrino in addition to three active neutrinos in order to introduce one extra mass-squared splitting (but -- at the same time -- not to change significantly the fit to solar, reactor, atmospheric and accelerator neutrino experiments).

While the 3+1 neutrino models with one light sterile neutrino are considered to be disfavored by 
present data [5], the 3+2 neutrino schemes with two light sterile neutrinos may provide a better 
description of current neutrino oscillations including the LSND effect [6]. However, the special 3+2 model of Ref. [7], where among three active neutrinos $\nu_e, \nu_\mu, \nu_\tau$ and two sterile neutrinos $ \nu_s, \nu_{s'}$ there are two maximally mixing pairs $\nu_\mu, \nu_\tau $ and
$\nu_s, \nu_{s'}$, does not meet these expectations. Also, the simplest 3+2 neutrino model considered in the present note is not better for explaining the LSND effect than the simplest 3+1 neutrino model [8].

For the experimental existence of two light sterile neutrinos (beside three generations of SM-active leptons and quarks) we argued some time ago in Refs. [9], where a Pauli principle working intrinsically within a generalized Dirac equation was introduced. In our simplest 3+2 neutrino model considered here, two light sterile neutrinos $\nu_s, \nu_{s'}$ mix very weakly with three active neutrinos $\nu_e, \nu_\mu, \nu_\tau$ and mutually do not mix at all.

The simplest 3+2 neutrino model is defined by the $5\times 5$ mixing matrix $U^{(5)} = \left( U^{(5)}_{\alpha i} \right)$ ($\alpha = e, \mu, \tau, s, s'$ and $i = 1,2,3,4,5$) of the form

\vspace{-0.2cm}

\begin{eqnarray}
U^{(5)} & = & U^{(5)}(12) U^{(5)}(14,25) \nonumber \\
 & = & \left( \begin{array}{ccccc} 
c_{12}c_{14} & s_{12}c_{25} & 0 & c_{12}s_{14} & s_{12}s_{25} \\ 
-\frac{1}{\sqrt2}s_{12}c_{14}\;\;\, & \frac{1}{\sqrt2}c_{12}c_{25} & \frac{1}{\sqrt2} & -\frac{1}{\sqrt2}   s_{12} s_{14}\;\;\, & \frac{1}{\sqrt2}c_{12}s_{25} \\ 
\frac{1}{\sqrt2}s_{12}c_{14} & -\frac{1}{\sqrt2}c_{12}c_{25}\;\;\, & \frac{1}{\sqrt2} & \frac{1}{\sqrt2} s_{12}s_{14} & -\frac{1}{\sqrt2}c_{12}s_{25}\;\;\, \\
-s_{14}\;\;\, & 0 & 0 & c_{14} & 0 \\ 0 & -s_{25}\;\;\, & 0 & 0 & c_{25} 
\end{array} \right)  \,,
\end{eqnarray}

\ni where 

\vspace{-0.2cm}

\begin{eqnarray}
U^{(5)}(12)\;\;\;\; & = & \left( \begin{array}{ccccc} 
c_{12} & s_{12} & 0 & 0 & 0 \\ 
\!-\frac{1}{\sqrt2}s_{12} & \frac{1}{\sqrt2}c_{12} & \frac{1}{\sqrt2}& 0 & 0 \\ 
\frac{1}{\sqrt2}s_{12} & \!\!-\frac{1}{\sqrt2}c_{12} & \frac{1}{\sqrt2} & 0 & 0 \\
0 & 0 & 0 & 1 & 0 \\ 
0 & 0 & 0 & 0 & 1 \end{array}\right) \,, \nonumber \\
U^{(5)}(14,25) & = & \;\;\left( \begin{array}{ccccc} 
c_{14} & 0 & 0 & s_{14} & 0 \\ 
0 & c_{25} & 0 & 0 & s_{25} \\ 0 & 0 & 1 & 0 & 0 \\
\!\!-s_{14} & 0 & 0 & c_{14} & 0 \\ 0 & \!\!-s_{25} & 0 & 0 & c_{25} \end{array} \right) \,.
\end{eqnarray}

\ni In this model, the extended unitary transformation (1) holds between the flavor neutrinos $\nu_e, \nu_\mu, \nu_\tau, \nu_s, \nu_{s'}$ and mass neutrinos $\nu_1, \nu_2, \nu_3, \nu_4, \nu_5$, where $U^{(3)}_{\alpha i}$ are replaced by $U^{(5)}_{\alpha i}$. Here, $c^2_{14} \gg s^2_{14}$ and $c^2_{25} \gg s^2_{25}$ guarantee the very weak mixing of two sterile neutrinos $\nu_s, \nu_{s'}$ with three active neutrinos $\nu_e, \nu_\mu, \nu_\tau$. Besides, there is no mixing between $\nu_s$ and $ \nu_{s'}$.

Then, making use of the extended neutrino oscillation formulae (3), where $U^{(3)}_{\alpha i}$ are replaced by $U^{(5)}_{\alpha i}$, we obtain (in the vacuum)

\begin{equation}
P(\nu_e \rightarrow \nu_e) \simeq 1 \!-\! 4c^2_{12}s^2_{12}\left(1 \!-\! s^2_{14} \!-\! s^2_{25}\right) 
\sin^2 x _{21} \!-\! 4c^2_{12}s^2_{14} \sin^2 x _{41} \!-\! 4s^2_{12}s^2_{25} \sin^2 x _{51}
\end{equation}

\ni when $x_{41}\simeq x_{42}$, $x_{51}\simeq x_{52}$, and

\begin{eqnarray}
P(\nu_\mu \rightarrow \nu_\mu) & \simeq & 1 - c^2_{12}s_{12}^2 \left(1 - s^2_{14} - s^2_{25}\right)
\sin^2 x _{21} - \left(1 - s^2_{12} s^2_{14} - c^2_{12} s^2_{25}\right) \sin^2 x _{32} \nonumber \\
& &  - 2s^2_{12}s^2_{14}\sin^2 x _{41} - 2c^2_{12} s^2_{25} \sin^2 x _{51} 
\end{eqnarray}

\ni when $x_{31}\simeq x_{32}$, $x_{41}\simeq x_{42}$, $x_{51}\simeq x_{52}$, as well as

\begin{eqnarray}
P(\bar{\nu}_\mu \rightarrow \bar{\nu}_e) & \simeq & 2c^2_{12}s^2_{12}\left(1 - s^2_{14} - s^2_{25}\right) \sin^2 x _{21} + 2c^2_{12}s^2_{12} s^2_{14} s^2_{25} \sin^2 x _{54} \nonumber \\
& & + 2c^2_{12} s^2_{12}\left(s^2_{14} - s^2_{25}\right) \left(s^2_{14} \sin^2 x _{41} - s^2_{25} 
\sin^2 x _{51} \right) 
\end{eqnarray}

\ni when $x_{41}\simeq x_{42}$, $x_{51}\simeq x_{52}$. In Eqs. (7) and (8), the terms $O(s^4_{14}) $, $O(s^4_{25})$ and $O(s^2_{14}s^2_{25})$ are neglected.

>From Eqs. (7) and (8) as well as (9) we get

\begin{eqnarray}
P(\nu_e \rightarrow \nu_e)_{\rm sol} & \simeq & 1 - 4c^2_{12}s^2_{12}\left(1\!-\! s^2_{14} \!-\! s^2_{25}\right) \sin^2 (x _{21})_{\rm sol} \!-\! 2\left(c^2_{12}s^2_{14} + s^2_{12} s^2_{25}\right) \nonumber \\ 
& =  & \left(1 \!-\! s^2_{14} \!-\! s^2_{25}\right) \left[1 \!-\! 4c^2_{12} s^2_{12}\sin^2 (x _{21})_{\rm sol}\right] \!-\! \left(c^2_{12}-s^2_{12}\right) \left(s^2_{14} \!-\! s^2_{25} \right) 
\end{eqnarray}

\ni when $x_{21} \ll x_{41} < x_{51}$ with $(x _{21})_{\rm sol} = O(\pi/2)$ or 

\begin{equation}
P(\bar{\nu}_e \rightarrow \bar{\nu}_e)_{\rm Chooz} \simeq 1 - 2\left(c^2_{12}s^2_{14} + s^2_{12} s^2_{25} \right) \sim 1\; ({\rm experimentally})
\end{equation}

\ni when $x_{21} \ll x_{41} < x_{51}$ with $(x _{31})_{\rm Chooz} \simeq (x _{31})_{\rm atm} = O(\pi/2)$, and

\begin{equation}
P(\nu_\mu \rightarrow \nu_\mu)_{\rm atm}  \simeq  \left(1 - s^2_{12}s^2_{14} - c^2_{12} s^2_{25} \right) \left[1- \sin^2 (x_{32})_{\rm atm}\right] 
\end{equation}

\ni when $x_{21} \ll x_{31} \ll x_{41} < x_{51}$ with $(x _{31})_{\rm atm} = O(\pi/2)$, as well as

\begin{equation}
P(\bar{\nu}_\mu \rightarrow \bar{\nu}_e)_{\rm LSND}  \simeq  2c^2_{12} s^2_{12}\left[ s^2_{14} s^2_{25} \sin^2 (x_{54})_{\rm LSND} + \frac{1}{2}\left( s^2_{14} - s^2_{25}\right)^2 \right] 
\end{equation}

\ni when $x_{21} \ll x_{54} \ll x_{41} < x_{51}$ with $(x _{54})_{\rm LSND} = O(\pi/2)$. In the symmetric case of $s^2_{14} \simeq s^2_{25}$, Eq. (13) reads

\begin{equation}
P(\bar{\nu}_\mu \rightarrow \bar{\nu}_e)_{\rm LSND} \simeq 2c^2_{12} s^2_{12} s^4_{14} \sin^2 (x_{54})_{\rm LSND} \,.
\end{equation}

If the nonzero LSND effect exists in the order $P(\bar{\nu}_\mu \!\rightarrow \!\bar{\nu}_e)_{\rm LSND} \!\sim\! (10^{-2}\;{\rm to}\;10^{-3})$ $ \times\sin^2 (x_{54})_{\rm LSND}$, the formula (14) valid in the case of $s^2_{14} \simeq s^2_{25}$ gives

\begin{equation} 
s^2_{14} \sim \left(\frac{10^{-2}\;{\rm to}\; 10^{-3}}{2c^2_{12}s^2_{12}}\right)^{1/2} \sim 0.15\; {\rm to} \;0.049 \,,
\end{equation}

\ni where $2c^2_{12} s^2_{12} \sim 0.42$ ($s^2_{12} \sim 0.30$). Then, in the case of $s^2_{12} \simeq s^2_{25}$

\begin{equation}
P(\nu_e \rightarrow \nu_e)_{\rm sol} \sim  [1 - (0.31 \;{\rm to}\; 0.098)][1 - 0.84 \sin^2 (x _{21})_{\rm sol}] 
\end{equation}

\ni or

\begin{equation} 
P(\bar{\nu}_e \rightarrow \bar{\nu}_e)_{\rm Chooz}  \sim  1 - (0.31 \;{\rm to}\; 0.098) \sim 1\; ({\rm experimentally}) \,,
\end{equation}

\ni and 

\begin{equation} 
P(\nu_\mu \rightarrow \nu_\mu)_{\rm atm} \sim [1 - (0.15 \;{\rm to}\; 0.049)][1 - \sin^2 (x _{32})_{\rm atm}]\,. 
\end{equation} 

\ni Here, $ \Delta m^2_{54} \equiv \Delta m^2_{\rm LSND} \sim 1\;{\rm eV}^2$ (say).

We can see that, in particular, the nonzero LSND effect of the order $P(\bar{\nu}_\mu \rightarrow \bar{\nu}_e) \sim 10^{-2} \sin^2(x_{54})_{\rm LSND}$ would imply an experimentally visible Chooz effect $P(\bar{\nu}_e \rightarrow \bar{\nu}_e) \sim 1- 0.31 < 1$, which is not observed. If its order was $P(\bar{\nu}_\mu \rightarrow \bar{\nu}_e) \sim 10^{-3} \sin^2 (x_{54})_{\rm LSND}$ the corresponding Chooz effect, $P(\bar{\nu}_e \rightarrow \bar{\nu}_e) \sim 1- 0.098 \sim 1$, would be nearly at the (Chooz) experimental edge (here, still $s_{13} = 0$).

In order to pass to the simplest 3+1 neutrino model [8], we can put $s_{25} \rightarrow 0$ and $ (x_{41})_{\rm LSND} =O(\pi/2)$, what gives

\begin{equation} 
P(\nu_e \rightarrow \nu_e)_{\rm sol} \simeq (1-s^2_{14})\left[1 - 4 c^2_{12}s^2_{12} \sin^2(x_{21})_{\rm sol}\right] - (c^2_{12}  - s^2_{12}) s^2_{14}  
\end{equation} 

\ni or

\begin{equation} 
P(\bar{\nu}_e \rightarrow \bar{\nu}_e)_{\rm Chooz}  \simeq 1 - 2 c^2_{12}s^2_{14} \sim 1 \;({\rm experimentally})\,, 
\end{equation} 

\ni and

\begin{equation} 
P(\nu_\mu \rightarrow \nu_\mu)_{\rm atm} \simeq (1- s^2_{12}s^2_{14})\left[1 - \sin^2(x_{32})_{\rm atm} \right] \,, 
\end{equation} 

\ni as well as

\begin{equation} 
P(\bar{\nu}_\mu \rightarrow \bar{\nu}_e)_{\rm LSND} \simeq 2 c^2_{12} s^2_{12} s^4_{14} \sin^2 (x_{41})_{\rm LSND} 
\end{equation} 

\ni (in Eqs (19), (20) and (21) the terms $O(s^4_{14})$ are neglected). Then, as before, $s^2_{14} \sim 0.15$ to 0.049 for the nonzero LSND effect with the amplitude of the order $O(10^{-2}\;{\rm to} \; 10^{-3})$. Here, $ \Delta m^2_{41} \equiv \Delta m^2_{\rm LSND} \sim 1\;{\rm eV}^2$ (say). Thus, the LSND effect is here the same as in our simplest 3+2 neutrino model with $s^2_{14} \simeq s^2_{25}$ and $(x_{54})_{\rm LSND} =O(\pi/2)$. But there, $ \Delta m^2_{54} \equiv \Delta m^2_{\rm LSND} \sim 1\;{\rm eV}^2$ (say). In both cases, the present (pre-MiniBooNE) data do not exclude a small LSND effect with the amplitude of the order $O(10^{-3})$.

\vfill\eject

~~~~

\vspace{0.5cm}

{\centerline{\bf References}}

\vspace{0.5cm}

{\everypar={\hangindent=0.7truecm}
\parindent=0pt\frenchspacing

{\everypar={\hangindent=0.7truecm}
\parindent=0pt\frenchspacing

[1]~For a review {\it cf.} S.~Pascoli, S.T.~Petcov and T. Schwetz, {\tt hep--ph/0505226}.

\vspace{0.2cm}

[2]~M. Apollonio {\it et al.} (Chooz Collaboration), {\it Eur. Phys. J.} {\bf C 27}, 331 (2003).

\vspace{0.2cm}

[3]~C. Athanassopoulos {\it et al.} (LSND Collaboration), {\it Phys. Rev. Lett.} {\bf 77}, 3082 (1996); {\it Phys. Rev. } {\bf C 58}, 2489 (1998); A. Aguilar {\it et al.}, {\it Phys. Rev.} {\bf D 64}, 112007 (2001).

\vspace{0.2cm}

[4]~A.O. Bazarko {\it et al.} (MiniBooNE Collaboration),  {\tt hep--ex/9906003}.

\vspace{0.2cm}

[5]~{\it Cf. e.g.} M. Maltoni, T. Schwetz, M.A.~Tortola and J.W. Valle, {\it Nucl. Phys.} {\bf B 643}, 321 (2002).

\vspace{0.2cm}

[6]~M. Sorel, J. Conrad and M. Shaevitz, {\tt hep--ph/0305255}.

\vspace{0.2cm}

[7]~W. Kr\'{o}likowski, {\it Acta Phys. Pol.} {\bf B 35}, 1675 (2004) [{\tt hep--ph/0402183}].

\vspace{0.2cm}

[8]~Ref. [7], Appendix A.

\vspace{0.2cm}

[9]~~W. Kr\'{o}likowski, in {\it Proc. of the 12th  Max Born Symposium, Wroc{\l}aw, Poland, 1998}, eds. A.~Borowiec {\it et al.}, Springer, 2000 [{\tt hep-ph/9808307}]; {\it Acta Phys. Pol.} {\bf B 30}, 227 (1999) [{\tt hep-ph/9808207}], Appendix; {\it Acta Phys. Pol.} {\bf B 31}, 1913 (2000) [{\tt hep-ph/0004222}]; {\it cf.} also {\it  Acta Phys. Pol.} {\bf B 32}, 2961 (2001) [{\tt hep-ph/0108157}]; {\it  Acta Phys. Pol.} {\bf B 33}, 2559 (2002) [{\tt hep-ph/0203107}]; {\tt hep-ph/0504256}. 

\vfill\eject

\end{document}